# Thermal Conductivity of Graphene Laminate: Making Plastic Thermally Conductive

H. Malekpour[1], K.-H. Chang[2], J.-C. Chen[2], C.-Y. Lu[2], D.L. Nika[1,3], K.S. Novoselov[4] and A.A. Balandin[1,*]

[1]*Nano-Device Laboratory and Phonon Optimized Engineered Materials – POEM – Center, University of California – Riverside, Riverside, California 92521 USA*

[2]*Bluestone Global Tech, Wappingers Falls, New York USA*

[3]*Department of Physics and Engineering, Moldova State University, Chisinau, MD-2009, Republic of Moldova*

[4]*School of Physics and Astronomy, University of Manchester, Manchester, UK*

**Abstract**

We have investigated thermal conductivity of graphene laminate films deposited on polyethylene terephthalate substrates. Two types of graphene laminate were studied – as deposited and compressed – in order to determine the physical parameters affecting the heat conduction the most. The measurements were performed using the optothermal Raman technique and a set of suspended samples with the graphene laminate thickness from 9 µm to 44 µm. The thermal conductivity of graphene laminate was found to be in the range from 40 W/mK to 90 W/mK at room temperature. It was found unexpectedly that the average size and the alignment of graphene flakes are more important parameters defining the heat conduction than the mass density of the graphene laminate. The thermal conductivity scales up linearly with the average graphene flake size in both uncompressed and compressed laminates. The compressed laminates have higher thermal conductivity for the same average flake size owing to better flake alignment. The possibility of up to 600× enhancement of the thermal conductivity of plastic materials by coating them with the thin graphene laminate films has important practical implications.

H. Malekpour, K.H. Chang, J.C. Chen, C.Y. Lu, D.L. Nika, K.S. Novoselov and A.A. Balandin (2014)

## CONTENTS IMAGE

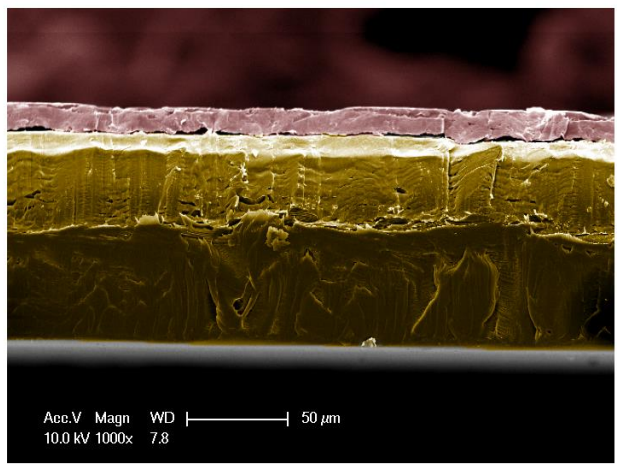
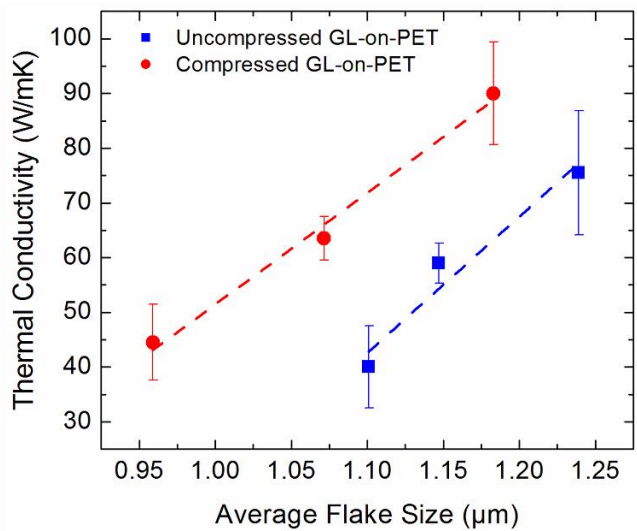

H. Malekpour, K.H. Chang, J.C. Chen, C.Y. Lu, D.L. Nika, K.S. Novoselov and A.A. Balandin (2014)

Graphene [1] is known to have exceptionally high intrinsic thermal conductivity [2-3]. The first measurements of the suspended graphene using the optothermal Raman technique revealed the thermal conductivity, $K$, values substantially exceeding that of the bulk graphite, which is $K$=2000 W/mK at room temperature (RT) [3]. The independent measurements with the optothermal method [4-5] and scanning thermal microscopy [6] confirmed this conclusion. Graphene can have higher $K$ than that of the basal planes of graphite despite similarities of the phonon dispersion and the crystal lattice inharmonicity. This intriguing fact was explained by the specifics of the long-wavelength phonon transport in two-dimensional (2-D) crystal lattices [3, 7-8]. The long-wavelength phonons in 2-D graphene have exceptionally long mean free path (MFP) limited by the size of the sample even if the thermal transport is diffusive [3]. In different terms, this means that the phonon Umklapp scattering alone is not sufficient for restoration of the thermal equilibrium in 2-D crystal lattice like in 3-D crystal lattices. The latter results in an anomalous dependence of the thermal conductivity of few-layer graphene (FLG) with the number of atomic planes in the samples [7]. The prediction of the logarithmic divergence of the intrinsic thermal conductivity of graphene with the size of the sample [8] has also recently found experimental confirmation [9]. Other interesting features of phonon thermal conduction in graphene include non-monotonic dependence on the graphene ribbon width [10] as well as strong isotope and point-defect scattering effects [11].

Excellent heat conduction properties of graphene coupled with graphene flexibility stimulated research of various composites where graphene and its derivatives play a role of filler materials [12-16]. Liquid phase exfoliated (LPE) graphene can be produced inexpensively and in large quantities, which makes graphene fillers practical. A proper mixture of LPE graphene and FLG flakes were shown to perform excellently as fillers in thermal interface materials (TIMs) [12-13] and thermal phase change materials (PCMs) [17]. Graphene is more promising as the filler than carbon nanotubes (CNTs) owing to its better thermal coupling to the matrix materials and substantially lower cost of LPE graphene. The loading fractions of graphene in composites are relatively low, e.g., up to 10% in TIMs [12] and up to 20% in PCMs [17]. Graphene and FLG flakes have also been used in metal templates for increased heat conduction [18]. In addition, it

H. Malekpour, K.H. Chang, J.C. Chen, C.Y. Lu, D.L. Nika, K.S. Novoselov and A.A. Balandin (2014)

has been shown that chemical vapor deposition (CVD) graphene coating can increase thermal conductivity of Cu films owing to the Cu grain size growth during CVD process [19].

A different type of graphene-based material with possible thermal coating applications is graphene laminate (GL) [20]. The chemically derived graphene and FLG flakes in GL layers are closely packed in overlapping structure. It is common to deposit GL films on polyethylene terephthalate (PET) substrates. The "sprayed on" GL films can be roll compressed. Given the growing practical needs in the thermally conductive coatings for various plastic materials it is interesting to study thermal properties of graphene laminates. The physics of heat conduction in such materials is non-trivial given the random nature of graphene flakes overlapping regions, a large distribution of the flake sizes and thicknesses as well as presence of defects and disorder. The knowledge of thermal properties of GL layers and understanding materials parameters that limit heat propagation can help in optimizing GL thermal coatings for practical applications.

In this paper we investigate the thermal conductivity of GL-on-PET using a set of as deposited and compressed samples with different mass densities and GL thickness ranging from ~9 μm to 44 μm. The measurements were performed using the optothermal Raman technique [2-3]. The optothermal technique originally introduced for μm-scale suspended graphene samples [2] has been extended to macroscopic suspended films. The rest of the paper is organized as follows. First, we describe the sample structure followed by the outline of thermal measurements and experimental results. The theory of heat conduction in FLG that compose GL is provided to assist in the experimental data analysis. Additional details of the sample preparation and thermal data extraction are given in the *Methods* section.

The graphene laminate layers were deposited on PET substrates. A subset of the samples was subjected to roll compression for this reason the samples were denoted "uncompressed" and "compressed". Additional details are provided in the Methods section. The widely available PET is a plastic material used for manufacturing of various containers. The thermal conductivity of

H. Malekpour, K.H. Chang, J.C. Chen, C.Y. Lu, D.L. Nika, K.S. Novoselov and A.A. Balandin (2014)

PET is extremely low in the range from 0.15 W/mK to 0.24 W/mK at RT. The low K value of PET limits its applications. The GL layer thickness was determined by the cross-sectional scanning electron microscopy (SEM) in the range from ~9 µm to 44 µm. Due to the thickness non-uniformity as average value among several locations was used in the analysis. The obtained GL layers had the mass density varying from 1.0 g/cm$^3$ to 1.9 g/cm$^3$. The electrical resistance was measured to be within the 1.8 Ω/□ - 6.1 Ω/□ interval. In Figure 1 we present cross-sectional SEM of the GL-on-PET where the PET substrate and GL layer can be easily distinguished.

The laminate is made of overlapping single layer graphene and FLG flakes with different size and shape. For quantitative analysis of thermal properties one needs accurate statistical data on the average flake size. This task is uneasy owing to the large size and shape variations. Figure 2 (a-b) shows representative SEM micrographs of the uncompressed and compressed GL-on-PET. We have performed extensive top view SEM studies to determine an average flake size *D*, which was defined as an average of the minimum and maximum diameter of each flake. More than hundred flakes have been taken into consideration for each sample in order to accumulate sufficient statistics for accurate calculation of *D*. Figure 3 shows the convergence of average flake size *D* to its apparent value for each sample. One can see that after ~50 flakes are included in the analysis the average size saturates to a particular well defined value. Table I provides nomenclature of the samples and their corresponding flake sizes.

We used non-contact optothermal Raman method for the thermal studies [3]. This is a direct steady-state measurement technique, which determines thermal conductivity directly without the need to calculate it from the thermal diffusivity data. In this technique, originally used for measuring the thermal properties of graphene [2], the micro-Raman spectrometer is used as thermometer to determine the local temperature rise. The Raman excitation laser is also used as a heater. The measurement procedure involved two steps: the calibration measurement and the power dependent Raman measurement. The micro-Raman spectroscopy (Renishaw In Via) was performed with 488 nm excitation laser and power level varied from 1 mW to 10 mW. The GL-on-PET samples were cut into the rectangular ribbon shapes (3 cm in length and 1 mm in width)

H. Malekpour, K.H. Chang, J.C. Chen, C.Y. Lu, D.L. Nika, K.S. Novoselov and A.A. Balandin (2014)

and suspended across a specially designed sample holder (see Figure 4). Massive aluminum pads – clamps served as ideal heat sinks and ensured good thermal contact with GL layers.

The calibration measurement was performed in the cold-hot cell (LINKAM THMS-600) with the temperature of the sample controlled externally. The equipment used for this measurement was capable of controlling temperature from -196°C up to 600°C with temperature stability below 0.1°C. Low excitation power of the Raman laser (~1 mW) was used to avoid the local laser induced heating. Since the low excitation power levels degrades the signal-to-noise (S/N) ratio we increased the exposure time to up to 10 second to achieve acceptable S/N ratio. The measurements were repeated for three times for each temperature to provide the average value. The calibration measurements give the Raman G peak position as a function of the temperature of the sample. Figure 5 shows the G peak spectral position as a function of temperature, $T$, for the interval from 20$^{o}$C to 200$^{o}$C for two uncompressed GL-on-PET samples. One can notice an excellent linear fitting over the examined temperature range and close values of the slope of the lines for two similar samples. The slope determines the temperature coefficient of G Raman peak for these given samples to be $\chi_G \approx -1.9 \times 10^{-2}$ cm$^{-1}$/K. It should be remembered that the $\chi_G$ value depends not only on the sample properties but also on the temperature range for which it was extracted.

The second part of the measurement is recording of the Raman G peak position of the suspended GL-on-PET sample (see Figure 4) as a function of the increasing excitation laser power. The power on the sample surface was measured by replacing the sample with the power meter (OPHIR). The absorbed power was determined by placing the power meter in the trench of the sample holder under the suspended GL-on-PET ribbon. The samples with various thickness of the GL layer were used to clarify the effects of the thickness. The results indicated that most of the power is absorbed by the sample and only a small portion (below 1%) is not absorbed owing to the leakage of the laser beam from the sample sides. The increase in the absorbed laser power, $\Delta P$, causes local heating which results in the red shift of the Raman G peak $\Delta \omega$. Figure 6 shows the measured G peak shift as the function of the absorbed power for the uncompressed and compressed samples. One can notice that two samples with somewhat different microstructure (compressed vs. uncompressed) have different temperature rise in response to the same heating power. The slope $\Delta\omega/\Delta P$ was measured to be -0.2451, -0.2255, -0.1521, -0.1776, -0.1766 and -0.1739 cm$^{-1}$/mW for GL-on-PET samples No. 1 to 6, respectively. Knowing the sample geometry and temperature rise, $\Delta T = \chi_G^{-1} \Delta\omega$, in response to the absorbed power, $\Delta P$, one can determine the thermal conductivity $K$ by solving the heat diffusion equation numerically. The details of the $K$ extraction procedure are given in the *Methods* section.

The measured RT thermal conductivity for different uncompressed and compressed GL-on-PET sample is shown in Figure 7 and summarized in Table II. There are a few interesting points to note. The overall thermal conductivity values for GL-on-PET are rather high $K \approx 40$ W/mK – 90 W/mK. Considering that PET and related plastic materials have $K \approx 0.15$ W/mK - 0.24 W/mK, coating PET with graphene laminate increases the thermal conductivity by more than two orders of magnitude (×170 - ×600). The measured data indicates that one can achieve high thermal conductivity in both compressed and uncompressed samples. The PET substrates had little if any effects on the heat transfer. No dependence on the GL layer thickness was observed. One can see from Figure 7 that $K$ for uncompressed and compressed samples scales linearly with the average flake size $D$. We did not found direct correlation of $K$ with the mass density of the samples. The overall $K$ values and $K$ dependence on the flake size indicate that the heat conduction in GL layers is limited by the flake boundaries rather than intrinsic properties of the graphene and FLG. Compression results in better alignment of the flakes or their closer attachment, which results at higher $K$ value for each size of the flake. This conclusion is supported by the analysis of the top view SEM images, which suggests that there are fewer vertically oriented flakes in the compressed samples. The misaligned flakes appear as bright white spots (see Figure 2).

Our finding that high $K$ can be achieved in both uncompressed and compressed GL-on-PET has practical implications. It suggests that graphene coating, without roll compression or other processing steps, can be beneficial for improving heat conduction properties of plastics. New applications of plastic materials, e.g. packaging or housing of electronic components, require

H. Malekpour, K.H. Chang, J.C. Chen, C.Y. Lu, D.L. Nika, K.S. Novoselov and A.A. Balandin (2014)

higher thermal conductivity. The latter is related to the increasing dissipated heat densities for modern electronics and optoelectronics. In this sense, graphene laminates may have potential as thermal coatings.

The in-plane heat conduction in GL-on-PET is defined by the thermal conductivity of individual FLG flakes and strengths of their attachment to each other. It is difficult to theoretically describe the thermal conductivity of graphene laminate due to uncertainty of such parameters as the strength of the coupling of the individual flakes and their mutual orientation. In order to gain insight into the heat conduction in GL-on-PET we modify the formulas derived by some of us for graphite thin films [10, 21]. The specifics of graphene laminates enter the model via characteristic dimensions of the flakes and defect concentrations. The thermal conductivity for the basal plane in graphene laminates can be written as [10, 22-23]:

$$K = K_{xx} = \frac{1}{L_x L_y L_z} \sum_{s,\vec{q}} \hbar \omega_s(\vec{q}) \tau(\omega_s(\vec{q})) \upsilon_{x,s} \upsilon_{x,s} \frac{\partial N_0}{\partial T}, \quad (1)$$

where $\tau(\omega_s(\vec{q}))$ is the relaxation time for a phonon with the frequency $\omega_s(\vec{q})$ from the *s*-th acoustic phonon branch ($s = LA, TA, ZA$), $\vec{q}(q_\parallel, q_z)$ is the phonon wave vector, $\upsilon_{x,s}$ is the projection of phonon group velocity, $N_0$ is the Bose-Einstein distribution function, $T$ is the temperature and $L_x$, $L_y$, $L_z$ are the sizes of the sample. The notations *LA*, *TA* and *ZA* indicate the longitudinal acoustic, the transverse acoustic and the out-of-plane acoustic phonon branches, correspondingly. Following the approach of Ref. [10], we define the phonon transport in GL to be two-dimensional (2D) for phonons with frequencies $\omega_s > \omega_{c,s}$ and three-dimensional (3D) for phonons with $\omega_s \leq \omega_{c,s}$, where $\omega_c$ is a certain low-bound cutoff frequency. Approximating the actual equal-energy surfaces $\omega_s(\vec{q})$ with the cylindrical surfaces one can rewrite Eq. (1) for the 2D and 3D parts of the thermal conductivity in the following form [10]:

$$\begin{aligned} K^{3D} &\equiv \frac{\hbar^2}{4\pi^2 k_B T^2} \sum_{s=TA,LA,ZA} \frac{1}{\upsilon_s^\perp} \int_0^{\omega_{c,s}} [\omega_s^\parallel(q_\parallel)]^3 \tau(\omega_s^\parallel) \upsilon_s^\parallel(q_\parallel)) \frac{\exp(\hbar \omega_s^\parallel / k_B T)}{[\exp(\hbar \omega_s^\parallel / k_B T) - 1]^2} q_\parallel d\omega_s^\parallel, \\ K^{2D} &= \frac{\hbar^2}{4\pi^2 k_B T^2} \sum_{s=LA,TA,ZA} \frac{\omega_{c,s}}{\upsilon_s^\perp} \int_{\omega_{c,s}}^{\omega_{\max,s}} [\omega_s^\parallel(q_\parallel)]^2 \tau(\omega_s^\parallel) \upsilon_s^\parallel(q_\parallel) \frac{\exp(\hbar \omega_s^\parallel / k_B T)}{[\exp(\hbar \omega_s^\parallel / k_B T) - 1]^2} q_\parallel d\omega_s^\parallel, \end{aligned} \quad (2)$$

where $v_s^\perp = \omega_{c,s}/q_{z,\max}$ and $\omega_{c,s}$ is the phonon frequency of *s*-th branch at *A* – point of graphite Brillouin zone.

In our calculations we take into consideration three mechanisms of phonon scattering [21-24]: Umklapp scattering $\tau_U(\omega_s^{\|}) = M v_s^2 \omega_{\max,s} / (\gamma_s^2 k_B T [\omega_s^{\|}]^2)$, point – defect scattering $\tau_{pd}(\omega_s^{\|}) = 4 v_s^{\|} / (S_0 \Gamma q [\omega_s^{\|}]^2)$ and scattering on the flake boundaries $\tau_b(\omega_s^{\|}) = D/v_s^{\|}$, where $\gamma_{LA} = 2$, $\gamma_{TA} = 1$ and $\gamma_{ZA} = -1.5$ is the branch-dependent average Gruneisen parameters, $\omega_{max,s}$ is the maximal frequency of *s*-th branch, $S_0$ is the cross-section area per atom, *M* is the graphene unit cell mass and $\Gamma$ is the measure of the strength of the point-defect scattering determined from the typical defect densities in the given material. The total phonon relaxation time $\tau$ was calculated from the Matthiessen's rule: $1/\tau = 1/\tau_{pd} + 1/\tau_U + 1/\tau_b$. The boundary scattering from the FLG edges is assumed to be completely diffused. We estimated the low-bound value of $\Gamma$ for GL-on-PET from the energy-dispersive X-ray spectroscopy (EDS). The characteristic material composition from EDS is 92% - 94% of C, 5.7% - 6.5% of O, 0.34% of Na and 0.56% of S. The $\Gamma$ parameter calculated with this impurity composition is 0.02 – 0.03. It does not include vacancies and related structural defects, which also contribute to the phonon – point defect scattering. For this reason, in our calculations we used $\Gamma$ values in the range 0.05 – 0.2, which is rather typical for semiconductor or electrically insulating technologically important materials.

In Figure 8 (a-b) we show the dependence of the thermal conductivity $K = K^{3D} + K^{2D}$ on the temperature for different values of the average flake size *D* and different $\Gamma$. The experimental points are shown by the red and pink circles. In the range of experimentally observed values of *K*, our simulations reveal the weak dependence of *K* on the temperature. This behavior is similar to that observed in polycrystalline materials, where the phonon scatterings on crystalline grains dominate [24]. Increase of *D* or decrease of $\Gamma$ restores the strong dependence of *K* on the temperature (see Figure 8 (b)), which is typical for the crystalline semiconductors and graphene [25]. The calculated dependence of *K* on *D* is weaker than that obtained experimentally. We attribute this discrepancy to different orientation and coupling of flakes in experimental samples, which has not been taken into account in our model description. The experimental observation, confirmed theoretically, that the thermal conductivity of the composite increases with the increasing size of the graphene fillers is in line with the literature reports for carbon nanotubes and other carbon allotropes [26-27]. The overall values of the thermal conductivity of graphene laminate at RT (K~90 W/mK) are substantially lower than those (K~3000 W/mK) measured for large suspended graphene layers [3].The latter is explained by the fact that the thermal conductivity of the laminates is limited not by the lattice dynamics of the graphene flakes but by their size, attachment to each other and orientation with respect to the heat flux.

In conclusion, we investigated thermal conductivity of graphene laminate films on PET. The thermal conductivity of graphene laminate was found to be in the range from 40 W/mK to 90 W/mK at RT. It was determined that the average size and the alignment of graphene flakes are more important parameters defining the heat conduction than the mass density of the graphene laminate. The thermal conductivity scales up linearly with the average graphene flake size in both as deposited and compressed laminates. The compressed laminates have higher thermal conductivity for the same average flake size owing to better flake alignment. The possibility of more than *two orders-of-magnitude* enhancement of the thermal conductivity of plastic materials by coating them with thin graphene laminate can be used for improving thermal management of electronic and optoelectronic packaging.

**METHODS**

**Graphene Laminate Preparation:** An aqueous dispersion of graphene nanoflakes, Grat-Ink 101N provided by Bluestone Global Tech, was used as the coating ink in this study. It was determined that less than 1-wt % of non-ionic polymer-type surfactants contained in the ink. The presence of the surfactants improved the dispersion of graphene flakes helping in deposition of a uniform film. A conventional PET film was used as a substrate. The graphene ink coated on PET by a laboratory slit coater (SECM02, Shining Energy, Taiwan), was dried at $80^oC$ for 10 minutes to form GL-on-PET samples. A compression roller (SERP02, Shining Energy, Taiwan) was further used to obtained compressed samples. The sheet resistance of the graphene laminate was measured by the 4-point probe (RM3000, Jandel, UK). A total of 10 measurements at different spots were carried out to obtain the average sheet resistance of each sample. The measurements of the mass density of the GL-on-PET samples and PET films were performed by stamping out with a disc stamper (diameter, 12 mm). To attain the average weight of graphene laminate, a total of 10 sets of weight for each sample were measured by a 5-digit analytical balance (XS-105DU, Mettler Toledo, US). With the weight, disc diameter and thickness of the coating, the density of graphene laminate was calculated.

**Analysis of the Flake Size and Orientation:** The average flake size was determined by using three intersecting lines from the different sides of the flake. The example of images used in the procedure is provided in the Supplementary Information. The fraction of the misaligned, e.g. vertically oriented flakes was determined following the following method. First, SEM images with low magnification have been utilized in order to cover a large area on the sample surface. Representative SEM images for the compressed and uncompressed samples are shown in the Supplementary Information. It is known that in SEM the areas possessing vertically oriented flakes absorb a higher fraction of the electron beam and thus appear as brighter spots in the image. The latter happens because the sharper areas on the sample surface produce higher electric field and, as a result, absorb a large fraction of the electron beam. In order to evaluate the fraction of the misaligned flakes vs. aligned flakes a special MATLAB code was written. The code calculates the fraction of bright pixels vs. total pixels for each SEM image. Each pixel is assigned a number from 0 to 255 where 0 corresponds to the darkest black and 255 to the lightest white. The code detects the fraction of pixels possessing the value from 230 to 255 which is defined as misaligned flakes. The results of the tests indicated that the uncompressed samples possessed a larger fraction of the misaligned flakes than the compressed samples.

**Thermal Conductivity Extraction Procedure:** In order to extract the thermal conductivity we solved the Fourier's equation for the specific sample geometry. Since the thickness of the GL layer is significantly large (~9 μm – 44 μm) the heat diffusion equation has to be solved for 3-D structure. We used COMSOL software package for numerical solution of the equation with proper boundary conditions. The laser spot heat source was assumed to have the Gaussian distribution of the power, *P(x, y, z)*, through the sample given as

H. Malekpour, K.H. Chang, J.C. Chen, C.Y. Lu, D.L. Nika, K.S. Novoselov and A.A. Balandin (2014)$$P(x,y,z) = \frac{P_{tot}}{0.5\sqrt{(2\pi\sigma)^3}} \exp(-\frac{x^2+y^2+z^2}{2\sigma^2}), \tag{M1}$$

where, $P_{tot}$ is the total absorbed power by the sample and $\sigma$ is the standard deviation of the Gaussian distribution function defined from the laser spot size. The full-width half maximum (FWHM) occurs at 0.5 µm which taken as the radius of the laser source. The two ends of the suspended GL-on-PET ribbon are attached to the heat sinks, which are modeled as being at RT. All other boundaries are defined as insulated from the environment, which means that the temperature gradient across the boundary is set to zero:

$$\vec{n}(k\nabla T) = 0. \tag{M2}$$

The heat diffusion equation is solved via the iteration procedure. We enter the total power and the thermal conductivity as the inputs to the equation and determine the temperature distribution as the result of simulations. The simulated temperature rise is compare with the measured temperature in the laser spot. The thermal conductivity is adjusted to higher or lower value based on the comparison. The task is simplified by introducing the slope parameter

$$\theta = \frac{\partial \omega}{\partial P} = \chi \frac{\partial T}{\partial P}. \tag{M3}$$

The simulated plot of $K$ vs. $\theta$ gives the actual value of thermal conductivity K for the measured value of the slope $\theta$. An example of the plot is shown in the *Supplementary Information*.

*Acknowledgements*

The work at UC Riverside was supported, in part, by the National Science Foundation (NSF) project ECCS 1307671 on engineering the thermal properties of graphene, by DARPA Defense Microelectronics Activity (DMEA) under agreement number H94003-10-2-1003, and by STARnet Center for Function Accelerated nanoMaterial Engineering (FAME) – Semiconductor Research Corporation (SRC) program sponsored by MARCO and DARPA.

*Author Contributions*

A.A.B. led the thermal data analysis and wrote the manuscript; K.S.N. coordinated the project, contributed to data analysis and manuscript preparation; K.H.C., J.C.C. and C.Y.L. prepared the samples; H.M. performed material characterization and thermal measurements; D.L.N contributed to the theory of heat conduction and data analysis.

*Author Information*

The authors declare no competing financial interests. Correspondence and requests for materials should be addressed to (A.A.B.) Alexander.Balandin@ucr.edu and (K.S.N) Konstantin.Novoselov@manchester.ac.uk

*Supporting Information Available:* The supporting information provides additional cross-sectional and top-view SEM of GL-on-PET, thermal conductivity measurements data. This material is available free of charge via the Internet at http://pubs.acs.org

H. Malekpour, K.H. Chang, J.C. Chen, C.Y. Lu, D.L. Nika, K.S. Novoselov and A.A. Balandin (2014)

## FIGURE CAPTIONS

**Figure 1:** Cross-sectional SEM images of the (a) uncompressed (sample #1) and (b) compressed (sample #5) GL-on-PET. The pseudo colors are used to indicate the graphene laminate (burgundy) and PET (yellow) layers. The graphene laminate layer of the uncompressed sample is ~44-µm thick while the PET substrate is ~110-µm thick in the uncompressed GL-on-PET. The laminate thickness variation is clearly seen from the micrograph.

**Figure 2:** Top-view SEM image of the (a) uncompressed (sample #3) and (b) compressed (sample #4) GL-on-PET. Graphene laminate consists of the overlapping layers of graphene and FLG flakes with arbitrary shapes and random in-plane orientation. Although most of the flakes are aligned along the PET substrate some of the flakes reveal vertical orientation seen as bright white areas on SEM images. Note that the number of the misaligned vertical flakes is substantially reduced in the compressed GL-on-PET samples.

**Figure 3:** Statistical analysis of the FLG flake size in GL-on-PET samples. The calculated average flake size is shown as a function of the number of flakes taken into account. The data is presented for the uncompressed (sample #1) and two compressed (samples #4 and #6) GL-on-PET. Note that the flake sizes converge to the asymptotic average values of 1.10, 1.18 and 0.96 after number of the accounted flakes exceeds about a hundred.

**Figure 4:** Optical image of the specially designed sample holder for optothermal Raman measurements with macroscopic thin films. The GL-on-PET sample under test (seen as gray ribbon) is suspended across a trench and fixed with two massive aluminum pads acting as the heat sinks. The ribbon is heated with the Raman laser in the middle. The experimental setup is a scaled up version of the original one used for the measurement of the thermal conductivity of graphene.

H. Malekpour, K.H. Chang, J.C. Chen, C.Y. Lu, D.L. Nika, K.S. Novoselov and A.A. Balandin (2014)

**Figure 5:** Raman G peak as a function of the sample temperature. The measurements were carried out under the low excitation power to avoid local heating while the temperature of the sample was controlled externally. Note an excellent liner fit for the examined temperature range. The obtained dependence is used as a calibration curve for the thermal measurement.

**Figure 6:** Raman G peak shift as a function of the laser power on the sample surface. The results are shown for the uncompressed (sample #1) and compressed (sample #4) GL-on-PET. The shift in G peak position with increasing power indicates the local temperature rise. The slope of these linear dependencies is used for the extraction of the thermal conductivity.

**Figure 7:** Thermal conductivity of GL-on-PET as a function of the average flake size D. The results are shown for the compressed (red circles) and uncompressed (blue rectangles) GL-on-PET samples. The dashed lines are to guide the eyes only. Note that the high thermal conductivity can be achieved in both uncompressed and compressed samples. For the same flake size D, the compressed samples have higher thermal conductivity than uncompressed ones owing to better flake alignment.

**Figure 8:** Calculated thermal conductivity as a function of temperature shown for different flake size $D$ and defect scattering strength $\Gamma$. Not that increasing $D$ or decreasing $\Gamma$ increases the thermal conductivity and strengthens its temperature dependence. The experimental data points are shown with the circles.

## References

[1] Geim, A. K.; Novoselov, K. S. *Nat. Mater.* **2007,** 6, 183-191.

[2] Balandin, A. A.; Ghosh, S.; Bao, W.; Calizo, I.; Teweldebrhan, D.; Miao, F.; Lau, C. N. *Nano Lett.* **2008**, 8, 902−907.

[3] Balandin, A. A. *Nat. Mater.* **2011**, 10, 569−581.

[4] Cai, W.; Moore, A. L.; Zhu, Y.; Li, X.; Chen, S.; Shi, L.; Ruoff, R. S. *Nano Lett*. **2010**, 10, 1645–1651.

[5] Chen, S. S.; Moore, A. L.; Cai, W. W.; Suk, J. W.; An, J. H.; Mishra, C.; Amos, C.; Magnuson, C. W.; Kang, J. Y.; Shi, L.; Ruoff, R. S. *ACS Nano* **2011**, 5 (1), 321−328.

[6] Yoon, K.; Hwang, G.; Chung, J.; Kim, H. G.; Kwon, O.; Kihm, K. D.; Lee, J. S. *Carbon* **2014**

[7] Ghosh, S.; Bao, W.; Nika, D. L.; Subrina, S.; Pokatilov, E. P.; Lau, C. N.; Balandin, A. A. *Nat. Mater*. **2010**, 9, 555–558.

[8] Nika, D. L.; Pokatilov, E. P.; Askerov, A. S.; Balandin, A. A. *Phys. Rev. B* **2009**, 79 (15), 155413.

[9] Xu, X.; Pereira, L. F. C.; Wang, Y.; Wu, J.; Zhang, K.; Zhao, X.; Bae, S.; Bui, C. T.; Xie, R.; Thong, J. T. L.; Hong, B. H.; Loh, K. P.; Donadio, D.; Li, B.; Özyilmaz, B. *Nature Communications* **2014**, 5, 3689.

[10] Nika, D.L.; Askerov, A.S.; Balandin, A.A. *Nano Lett.* **2012**, *12*, 3238-3244.

[11] Chen, S.; Wu, Q.; Mishra, C.; Kang, J.; Zhang, H.; Cho, K.; Cai, W.; Balandin, A. A.; Ruoff, R. S. *Nat. Mater*. **2012**, 11 (3), 203−207.

[12] Shahil, K. M. F.; Balandin, A. A. *Nano Lett.* **2012**, 12, 861−867.

[13] Goyal, V.; Balandin, A. A. *Appl. Phys. Lett*. **2012**, 100, 073113.

[14] Eda, G.; Chhowalla, M. *Nano Lett.* **2009**, 9, 814-818.

[15] Huang, X.; Yin, Z.; Wu, S.; Qi, X.; He, Q.; Zhang, Q.; Yan, Q.; Boey, F.; Zhang, H. *Small* **2011**, 7, 1876−1902.

[16] Wu, Z. S.; Ren, W. C.; Gao, L. B.; Zhao, J. P.; Chen, Z. P.; Liu, B. L.; Tang, D. M.; Yu, B.; Jiang, C. B.; Cheng, H. M. *ACS Nano* **2009**, 3, 411–417.

[17] Goli, P.; Legedza, S.; Dhar, A.; Salgado, R.; Renteria J.; Balandin A.A. *J. Power Sources* **2014,** 248, 37-43.

[18] Boden, A; Boerner, B; Kusch, P; Firkowska, I; Reich, S. *Nano Lett. ASAP* **2014**

[19] Goli, P; Ning, H; Li, X; Lu, C.Y; Novoselov, K.S; Balandin, A.A, *Nano Lett*. **2014**, 14, 1497−1503

[20] Novoselov, K. S.; Falko, V. I.; Colombo, L.; Gellert, P. R.; Schwab, M. G.; Kim, K. *Nature* **2012**, 490, 192−200.

[21] Nika, D.L.; Pokatilov, E.P.; Balandin, AA. *Physica Status Solidi* B, **2011**, 248, 2609.

[22] Klemens, P.G. *J. Wide Bandgap Mater.* **2000**, *7*, 332-339.

[23] Nika, D.L.; Balandin, A.A. *J. Phys.: Condens. Matter*. **2012,** 24, 233203.

[24] Klemens, P. G. *Int. J. Thermophys.* **1994**,15 (6), 1345−1351.

[25] Morelli, D. T.; Slack, G. A. High Lattice Thermal Conductivity Solids. In *High Thermal Conductivity Materials*; Shinde, S. L., Goela, J. S., Eds.; Springer-Verlag: New York, 2006; p 37-68.

[26] Gonnet, P.; Liang, Z.; Choi, E. S.; Kadambala, R. S.; Zhang, C.; Brooks, J. S.; Wang, B.; Kramer, L. *Current Appl. Phys*. **2006**, 6, 119-122.

[27] Han, Z.; Fina, A. *Progress in Polymer Science* **2011**, 36, 914-944.

H. Malekpour, K.H. Chang, J.C. Chen, C.Y. Lu, D.L. Nika, K.S. Novoselov and A.A. Balandin (2014)

**Table I: Sample Nomenclature**

| GL-on-PET | Laminate Thickness [μm] | Average Flake Size [μm] | Note |
|---|---|---|---|
| 1 | 44 | 1.10 | Uncompressed |
| 2 | 14 | 1.15 | Uncompressed |
| 3 | 13 | 1.24 | Uncompressed |
| 4 | 9 | 1.18 | Compressed |
| 5 | 24 | 1.07 | Compressed |
| 6 | 30 | 0.96 | Compressed |

**Table II: Thermal Conductivity of GL-on-PET at RT**

| GL-on-PET | Average flake size [μm] | K [W/mK] | Note |
|---|---|---|---|
| 1 | 1.10 | 40±7.5 | Uncompressed |
| 2 | 1.15 | 59±3.6 | Uncompressed |
| 3 | 1.24 | 75.5±11.3 | Uncompressed |
| 4 | 1.18 | 90±9.4 | Compressed |
| 5 | 1.07 | 63.5±4.0 | Compressed |
| 6 | 0.96 | 44.5±6.9 | Compressed |

H. Malekpour, K.H. Chang, J.C. Chen, C.Y. Lu, D.L. Nika, K.S. Novoselov and A.A. Balandin (2014)

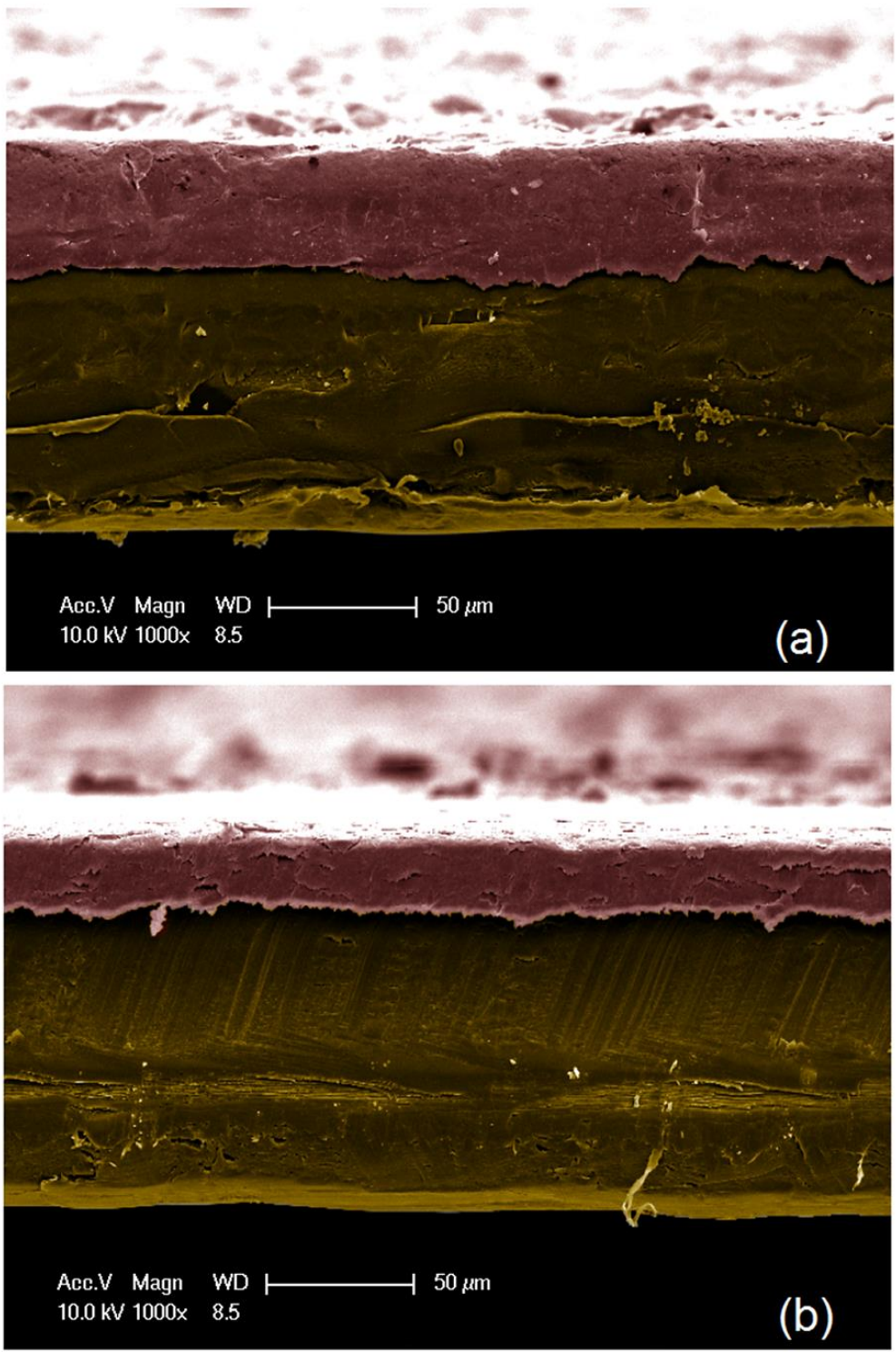

Figure 1

H. Malekpour, K.H. Chang, J.C. Chen, C.Y. Lu, D.L. Nika, K.S. Novoselov and A.A. Balandin (2014)

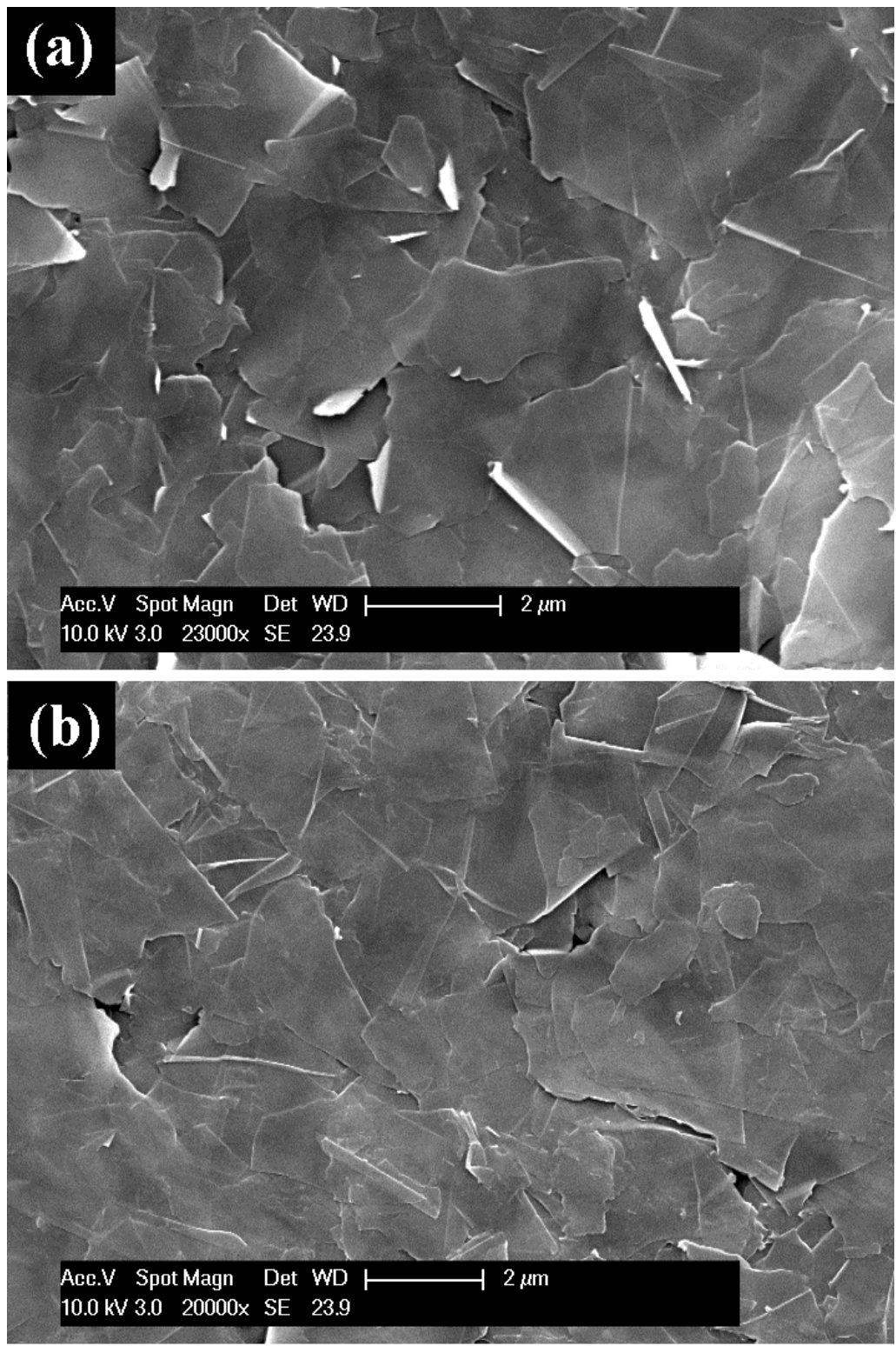

Figure 2

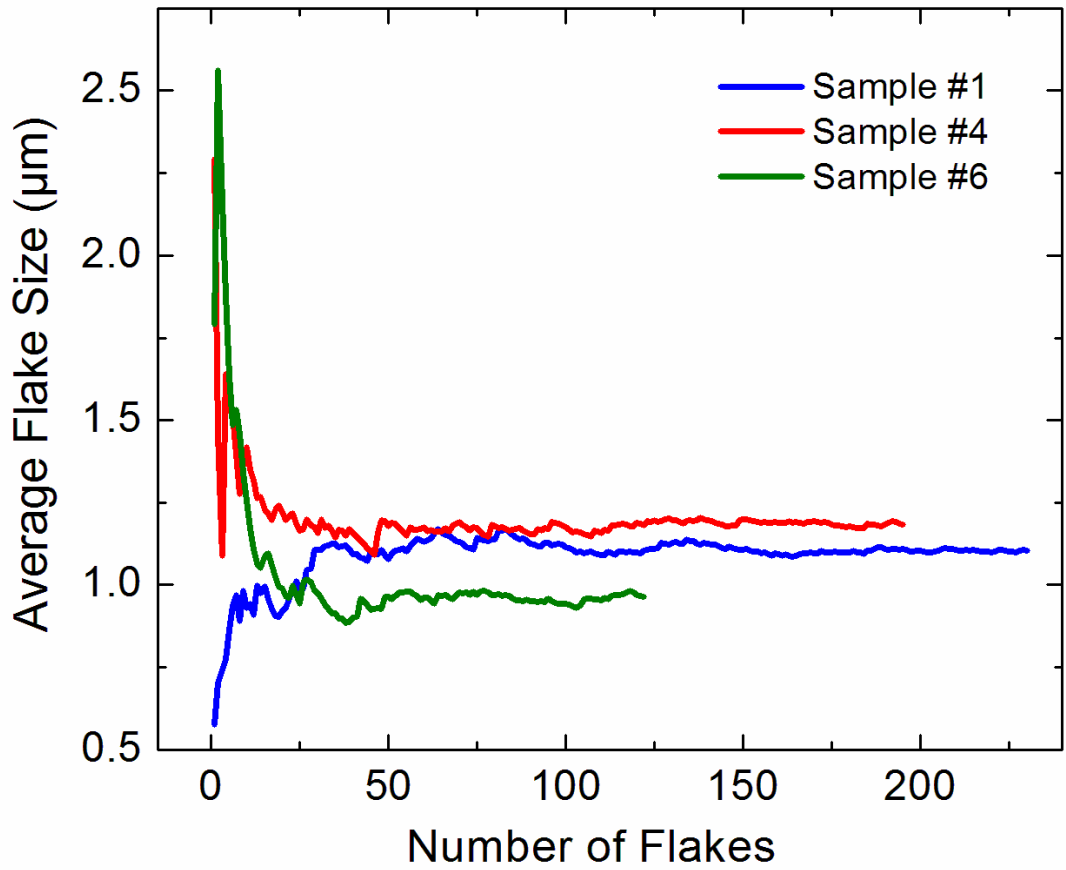

Figure 3

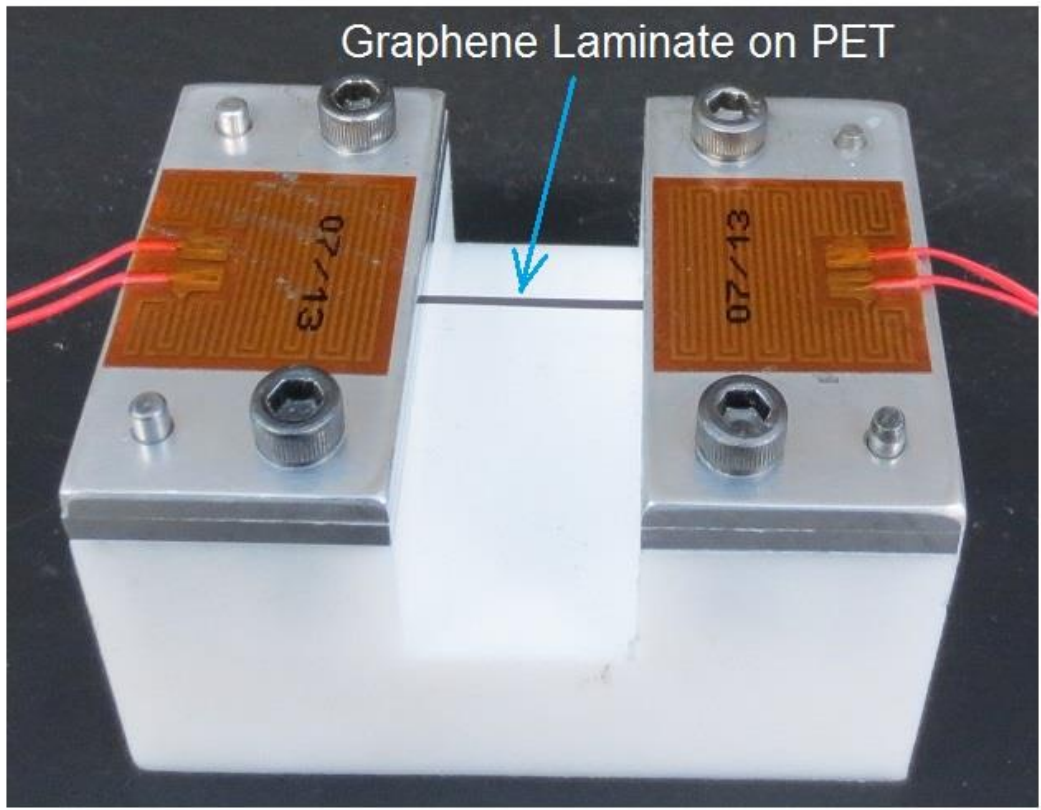

Figure 4

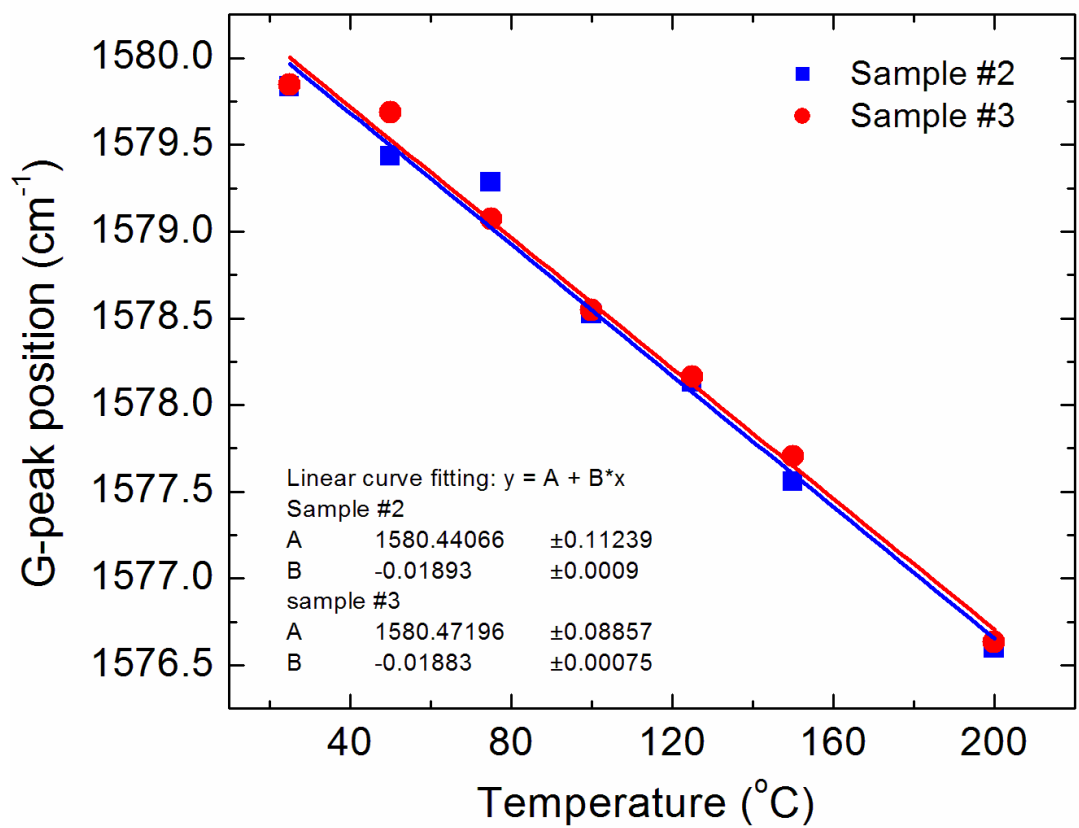

Figure 5

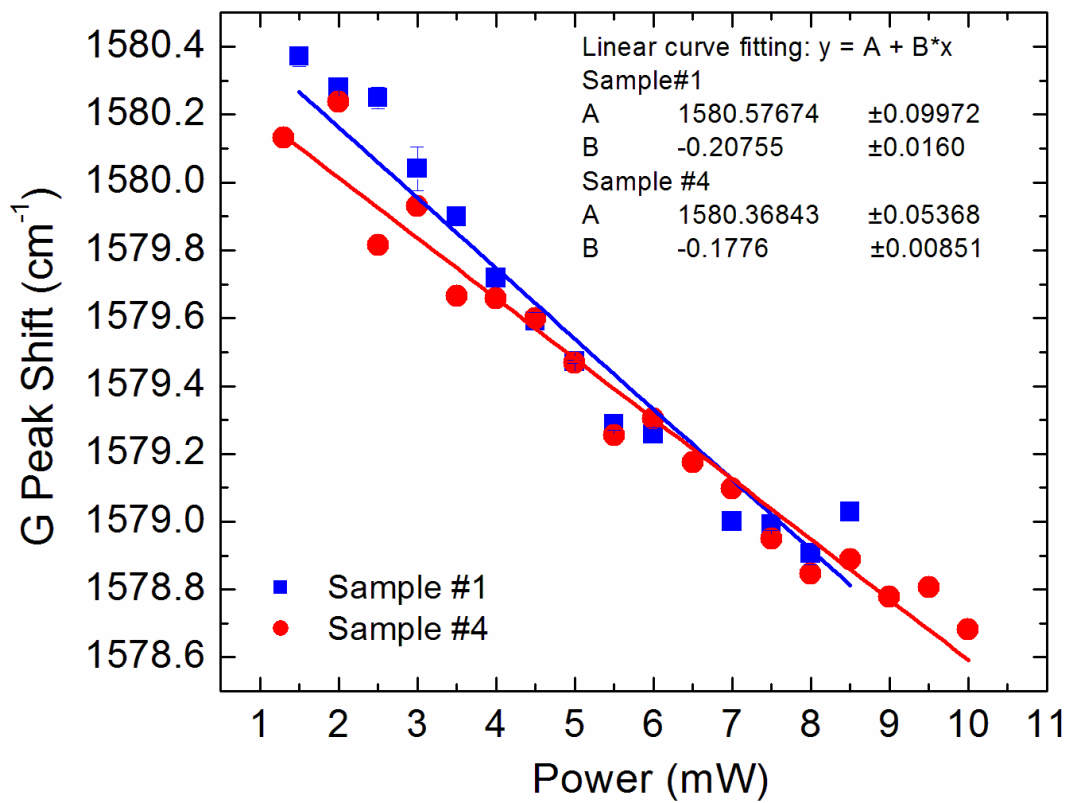

Figure 6

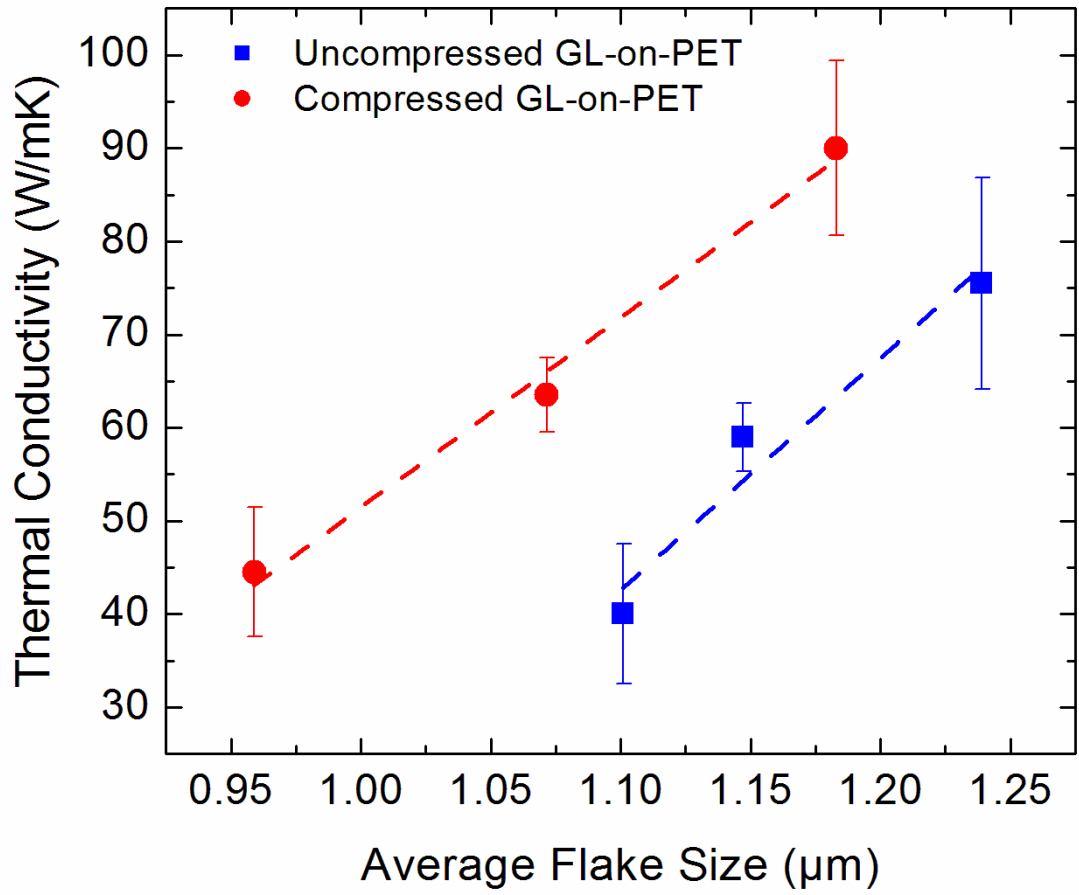

Figure 7

H. Malekpour, K.H. Chang, J.C. Chen, C.Y. Lu, D.L. Nika, K.S. Novoselov and A.A. Balandin (2014)

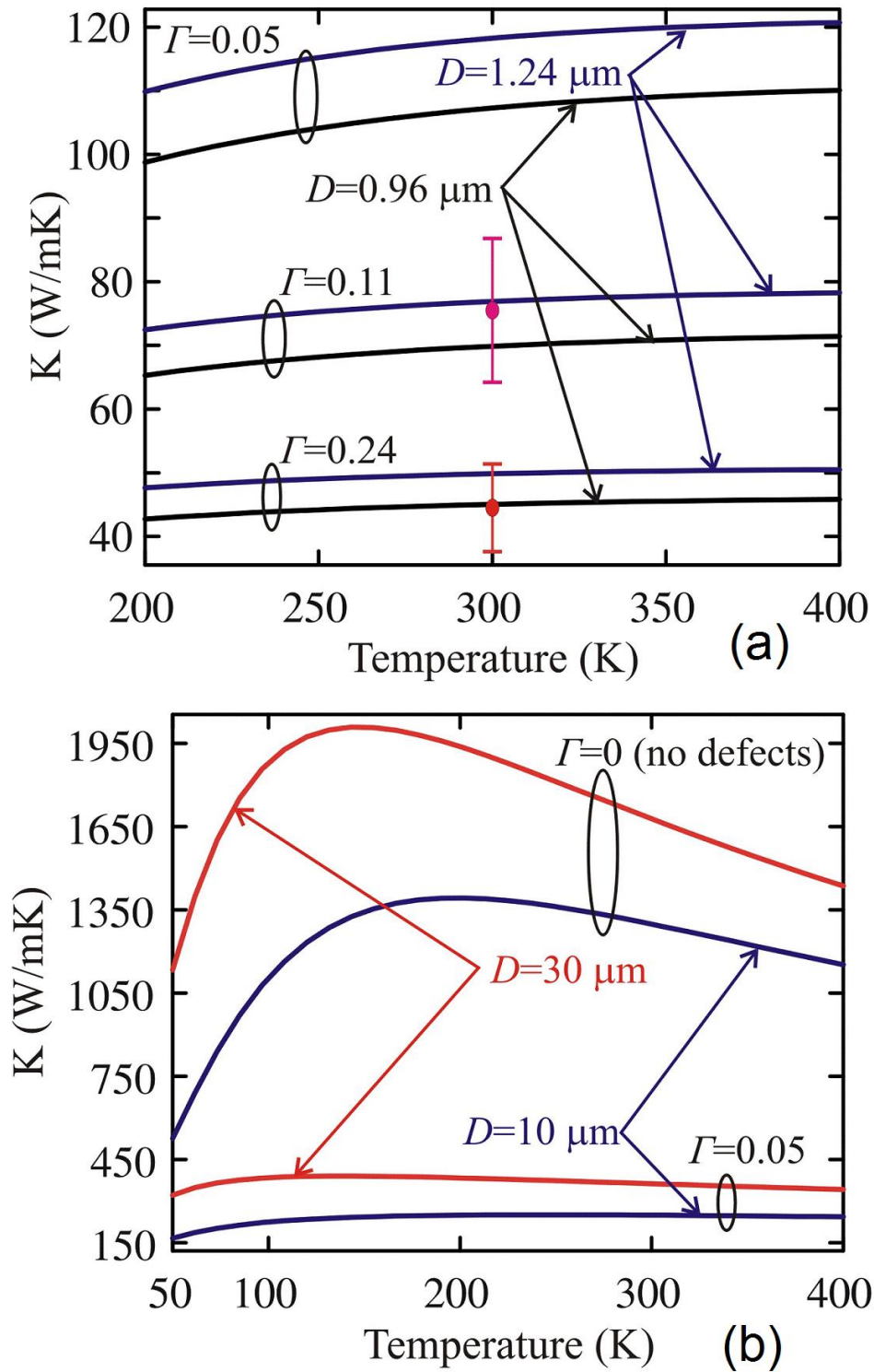

Figure 8